\documentstyle[12pt]{article}

\newcommand\noi{\noindent}
\newcommand\beq{\begin{equation}}
\newcommand\eeq{\end{equation}}
\newcommand\beqn{\begin{eqnarray}}
 \newcommand\eeqn{\end{eqnarray}}
\newcommand{\la}{\langle}
 \newcommand{\ra}{\rangle}
\relax

\begin{document}
\setlength{\baselineskip}{3.0ex}
\phantom{bla}

 \hspace*{9cm}{\large
 nucl-th/9606010}
 \vspace*{1.2cm}

\begin{center}
{ {\bf INTERPLAY OF INTERFERENCE EFFECTS}\\
 {\bf IN PRODUCTION OF $J/\Psi$ AND $\Psi'$ OFF NUCLEI}
\footnote{Invited talks presented by B.K. at 
the XXXI st Rencontres de Moriond,
''QCD and High-Energy Hadronic Interactions'' 
March 23-30, 1996, Les Arcs, France
and at PANIC'96, May 22-28, 1996, Williamsburg, USA}}\\ 
  \vspace{1cm}
{\bf J\"org H\"ufner}\\
Institut f\"ur Theor. Physik der Universit\"at\\
 Philosophenweg 19, 69120 Heidelberg,
Germany\\
\medskip
{\bf Boris Kopeliovich}\\
Max-Planck-Institut f\"ur
Kernphysik, 69029 Heidelberg,
Germany, and\\
Joint Institute for
 Nuclear Research, Dubna, 141980
Moscow Region, Russia\\
\vspace{3.5cm}

\vskip.5cm
{\bf Abstract}
\end{center}
\vbox{ \baselineskip 14pt
Our main observations are:

\medskip

\noi
(i) The effective path length of the $c\bar c$
wave packet, which is produced in the nucleus,
grows with the energy of the produced charmonium.
The variation is controlled by the coherence
length $l_c=2E_{\Psi}/M^2_{\Psi}$.\\
(ii) A colorless $c\bar c$ wave packet produced in
$pp$-interaction is a specific linear combination
of the $J/\Psi$ and $\Psi'$ states, and interacts
substantially weaker that any of its components.
The time evolution of this wave packet is
controlled by the formation length,
$l_f=2E_{\Psi}/(M^2_{\Psi'} - M^2_{\Psi})$.

\medskip

\noi
Exact formulas incorporating with these effects are derived.
The interplay of the two phenomena results in a
nontrivial energy- and $x_F$-dependence of the
nuclear suppression of the charmonium production
in proton-nucleus and nucleus-nucleus collisions
and explains some of the experimentally observed
effects.
}

\vspace*{3cm}
%
\setcounter{footnote}{0}
\newpage
\noi{\bf 1. Introduction}

\medskip

The experiments of producing a $\Psi$ (we use the
symbol $\Psi$ instead of $J/\Psi$) or a $\Psi'$
meson in a collision of a hadron $h$ with a
nucleus A, at energies $E_h$ of several hundreds
of $GeV$ have yielded a number of unexpected
results, of which we will recall the most
significant ones.  Since we want to limit
ourselves to nuclear effects and not absolute
cross sections, it is convenient to introduce the
nuclear suppression function
 \beq
 S^{hA}_{\Psi}(E,x_F) = {1\over A}
 \frac{d\sigma^{hA\to \Psi
 A}(E_h,x_F)/dx_F}
 {d\sigma^{hN\to \Psi A}(E_h,x_F)/dx_F}\ ,
\label{1a}
\eeq
which depends on the energy $E_h$ of the hadron,
and the Feynman variable $x_F$ of the $\Psi$.

The experimental results under discussion here
are:\\
(i) The value for $S^{pA}_{\Psi}(E_h,x_F)$ in the
interval $0 < x_F < x_0$ seems to depend on $E_h$,
the suppression factor being smaller for $E=800\
GeV$ $^{1)}$ than for $E=200\ GeV$ $^{2)}$.\\
(ii) For $x_F > 0$ in $pA$ collisions one has
nearly the same nuclear suppression for the $\Psi$
and $\Psi'$ mesons$^{1)}$.\\
(iii) $\Psi'$ turns out to be more suppressed than
$J/\Psi$ in nucleus-nucleus collisions,
$S^{AB}_{\Psi'} < S^{AB}_{\Psi}$ $^{3)}$.

\smallskip

In the present paper we discuss quantum
interference effects, which are not ad hoc
mechanisms (and do not need any unknown
parameters), but can explain, at least partially,
the above effects.\\

\bigskip

\noi{\bf 2. Glauber theory. The coherence length.}

\medskip

Nuclear suppression of charmonium depends on the
production mechanism even in the simplest case
of eikonal approximation.  Charmonium production
on a nucleon at high energy may be seen in the
lab.  frame as interaction of a fluctuation of the
projectile hadron, containing charm quarks, which
frees the charmonium.  We single out two types of
interaction:\\
a) Direct interaction of the $c\bar c$ projectile 
fluctuation with the target.  This
interaction must me sufficiently hard to resolve
the size of the $c\bar c$ pair in order to make it
colorless.\\
b) Interaction of the light spectator partons
accompanying the $c\bar c$ pair, freeing
the charmonium.  This interaction
can be soft i.e. have a large cross section at $x_1
\to 1$ $^{4)}$.

We assume hereafter the dominance of the direct
mechanism a), which restricts our consideration to
small values of $x_F$.

Since the charmonium production is a hard process,
a soft initial/final state interaction, which cannot
resolve the $c\bar c$ fluctuation, does not
produce any shadowing.	Charmonia produced on
different nucleons add up incoherently, since the
longitudinal momentum transfer is large, $q_L=
E_q(1-x_1)$, where $x_1=(x_F +
\sqrt{x_F^2+4M_{\Psi}^2/s})/2$.

There is also a possibility of an additional hard
scattering which frees the $c\bar c$ pair
''elastically'' in advance of the inelastic
interaction.  Namely, the projectile hadron can
experience a hard diffractive excitation with a
colorless exchange (Pomeron) in $t$-channel, which
puts the $c\bar c$ fluctuation on mass shall.  The
longitudinal momentum transfer to the target
nucleon may be small, provided that the energy is
high,
\beq q_c \approx
\frac{M^2_{\Psi}}{2E_{\Psi}}
\label{1}
\eeq
\noi
If $q_c \ll 1/R_A$, different nucleons contribute
coherently.

The excited hadron propagates through the
nucleus and produces the final charmonium with
energy $x_1E_h$ in another interaction.  Although
the hard diffractive cross section is quite small,
such a correction turns out to be very important, 
since it substantially increases the attenuation of the 
charmonium at high energy.

We skip the full expression for $S^{\gamma
A}_{\Psi}$, which is too lengthy and can be found
in $^{5,6)}$, where it was derived for the
first time.  That expression can be simplified
using smallness of $\sigma^{\Psi N}_{tot}\la T\ra
\ll 1$ (compare with $^{7)}$).
\beq
S^{hA}_{\Psi}(E_{\Psi}) \approx 1 -
{1\over 2}\ \sigma^{\Psi}_{in} \
\langle T \rangle
\left [1 + F_A^2(q_c) \right ]\ ,
\label{1b}
\eeq
\noi
where $\la T\ra$ is the mean nuclear thickness and 
the nuclear ''longitudinal formfactor'',
 \beq
 F^2_A(q_c) =
\frac{1}
 {A\langle T
 \rangle}\
 \int d^2b\
 \left |
\int_{-\infty}^{\infty} dz
 \rho_A(b,z)
 e^{iq_cz}\right |^2
\label{2}
\eeq
\noi
takes into account the phase shifts between the
waves produced at different points.

We conclude from (\ref{2}) that the nuclear
shadowing correction at high energy
($F^2_A(q_c)\to 1$) is twice as big as at low energy
($F^2_A(q_c)\to 0$).  This important result has a
natural space-time interpretation:  at high energy
the lifetime of the $c\bar c$ fluctuation of the
photon, $t_c=1/q_c$ (called coherence time or length), 
is long and the mean path of the $c\bar c$
pair in the nucleus is doubled compared to that at low
energy $^{8)}$.  Thus, $S^{hA}_{\Psi}$ decreases with
$x_F$.\\

\bigskip

 \noi{\bf 3. Beyond the Glauber model. The formation
length}

\medskip

In order to improve the eikonal Glauber
approximation one should take into account the off
diagonal diffractive rescatterings of the
charmonium in the nucleus.  We restrict our
consideration to a two-coupled-channel problem,
including $\Psi = {1\choose 0}$ and $\Psi' = 
{0\choose 1}$.  The initially produced $c\bar
c$ state $\Psi_0$ has its representation
 \beq
 |\Psi_0\rangle = \frac{1}{\sqrt{1+R^2}}
 {1\choose{R}}\ ,
\label{13a}
\eeq
The evolution of this state through the
nucleus can written in matrix representation,
 \beq
i\frac{d}{dz}{\alpha(z)\choose{\beta(z)}} =
 \widehat
U(b,z){\alpha(z)\choose{\beta(z)}}\ ,
\label{14a}
\eeq
 where
 \beq
 \widehat U = \left(\begin{array}
 {cc}q_1&0\\0&q_2
\end{array}\right) -
 \frac{i}{2}\ \sigma^{\Psi N}_{tot}\ \rho_A(b,z)
\left(\begin{array}
 {cc}1&\epsilon\\\epsilon&r \end{array}\right)\ ,
\label{15a}
\eeq
 with
 \beq
 \epsilon = \frac{\la\Psi'|\widehat f|\Psi\rangle}
{\la\Psi|\widehat f|\Psi\rangle}
\label{16a}
\eeq
 and
 \beq
 r = \frac{\la\Psi'|\widehat f|\Psi'\rangle}
{\la\Psi|\widehat f|\Psi\rangle} =
 \frac{\sigma^{\Psi' N}_{tot}}{\sigma^{\Psi
N}_{tot}}\ ,
\label{17a}
\eeq
$\widehat f$ is the operator of the $c\bar
c$-nucleon diffractive scattering amplitude.
$q_1$ and $q_2$ are the transferred longitudinal
momenta in photoproduction of $\Psi$ and $\Psi'$
respectively, as it is defined in eq.~(\ref{1}).
\beq
q_f = q_2 - q_1 =
\frac{M^2_{\Psi'}-M^2_{\Psi}}{2E_{\Psi}},\
\label{3}
\eeq
Exact solution is presented in $^{5)}$, however,
it is instructive to solve the equation
(\ref{14a}) to first order in $\sigma^{\Psi
N}_{tot}$, neglecting the effect of
coherence length considered above.  Then we find
for the nuclear suppression of the $\Psi$ and
$\Psi'$ states
 \beq
 S^{pA}_{\Psi}(E_{\Psi}) \approx
1-{1\over
2}\sigma^{\Psi N}_{in}
 \ \la T\rangle\ \left[1+\epsilon\
R\ F^2_A(q_f)\right]
\label {18a}
\eeq
 \beq
 S^{pA}_{\Psi'}(E_{\Psi}) \approx 1-
{1\over 2}\sigma^{\Psi' N}_{in}\
\la T\rangle\ \left[1+(\epsilon/r
R)\ F^2_A(q_f)\right]\ ,
\label {18b}
\eeq
We estimate matrix elements (\ref{16a}) -
(\ref{17a}) at $\epsilon = -\sqrt{2/3}$ and $r=7/3$ and
from experimental data $|R_{ex}| = 0.48 \pm 0.06$
$^{5)}$.  As opposite to the effect of the
coherence length discussed in the previous
section, the growth of the formation length leads
to the increase of $S^{pA}_{\Psi}$.  This is because
the produced initial state $|\Psi_0\rangle$ turns
out to be nearly an eigenstate of interaction,
provided that the parameters $\epsilon$, $r$ and $R$ have
values we estimated.  Such an eigenstate has the
absorption cross section smaller than any of its
components, $\Psi$ or $\Psi'$.  This explains the
growth of $S^{pA}_{\Psi}$ with $E_{\Psi}$, 
predicted by (\ref{18a}).

\begin{figure}[tbh]
\includegraphics{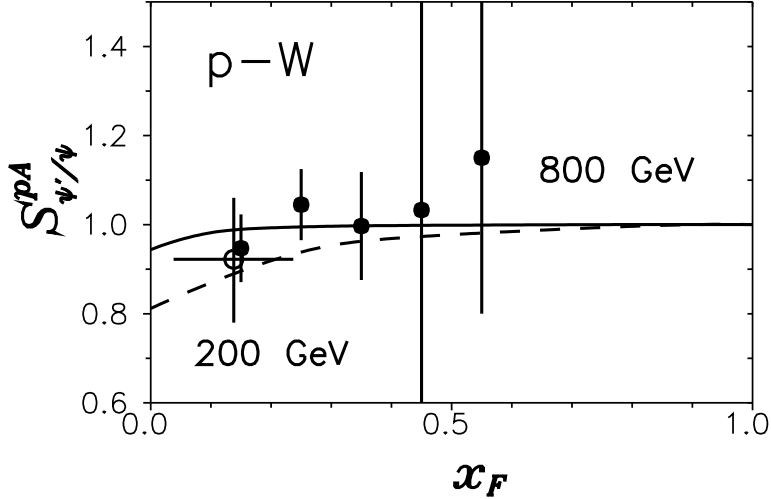}
\begin{center}
\vspace{6.5cm}
\parbox{13cm}
 {\caption[Delta]
{The relative nuclear suppression $S^{pA}_{\Psi'/\Psi}=
S^{pA}_{\Psi'}/S^{pA}_{\Psi}$ in $p-W$ collision at $800$ and
$200\ GeV$,
calculated in the two-coupled channel 
approach $^{5)}$ at in comparison with data $^{1,3)}$.}
\label{fig1}}
\end{center}
\end{figure}

As soon as the produced $c\bar c$ state
$|\Psi_0\ra$ is the eigen state, it does not
change its $\Psi$-$\Psi'$ content during
propagation through the nucleus, i.e.  the
relative yields of $\Psi'$ to $\Psi$ has no
$A$-dependence at high $E_{\Psi}$ as was observed 
in $^{1)}$.  This is
demonstrated in Fig.~\ref{fig1} versus $x_F$
$^{5)}$ in comparison with data
$^{1,3)}$.

\begin{figure}[tbh]
\includegraphics{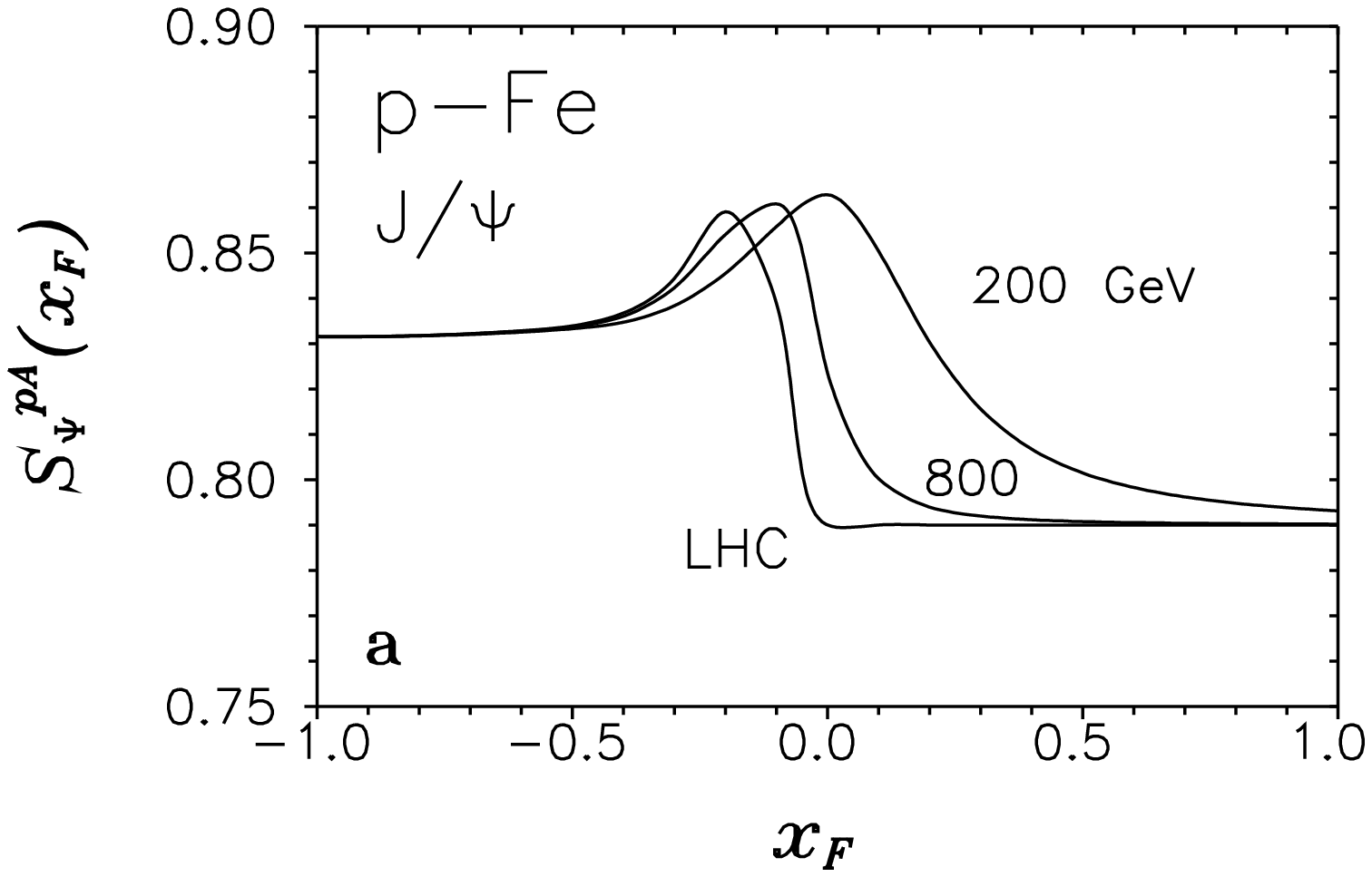}
 \includegraphics{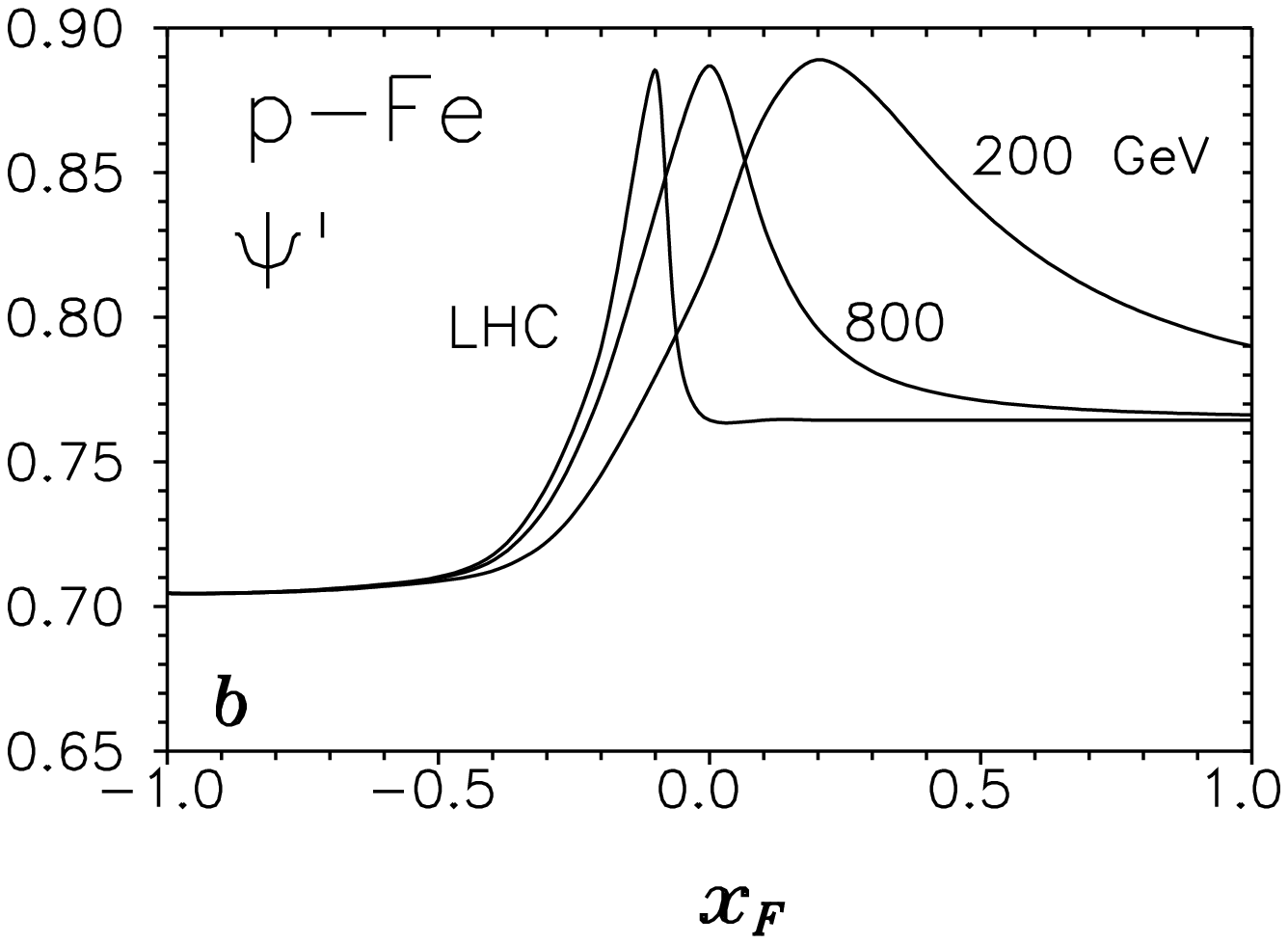}
\begin{center}
\vspace{6cm}
\parbox{13cm}
 {\caption[Delta]
{$x_F$-dependence of the nuclear suppression for
production of $\Psi$ ({\bf a}) and $\Psi'$ ({\bf
b}) in p-Fe collisions.  The curves show
predictions at the proton energies $200,\ 800\
GeV$ and in the energy range of RHIC - LHC.}
\label{fig2}}
\end{center}
\end{figure}

Now we can turn on the coherence length and
combine the two effects.  In the approximation of
small $\sigma_{tot}^{\Psi N}\la T\ra \ll 1$ we get
 \beq
S^{hA}_{\Psi}(E_{\Psi}) \approx
 1 - {1\over 2}\ \sigma^{\Psi}_{in}\
 \langle T
\rangle
 \left [1 + F_A^2(q_c)\right]\ \left[1 +
\epsilon\ R\ 
 F_A^2(q_f) \right ]
\label{5}
\eeq
 \beq
 S^{hA}_{\Psi'}(E_{\Psi}) \approx
 1 - {1\over 2}\
\sigma^{\Psi'}_{in}
 \langle T \rangle
 \left [1 + F_A^2((q_c)\right]\ \left[1 +
\frac{\epsilon}{r\ R}\ 
 F_A^2(q_f) \right ]
\label{6}
\eeq
Since the effects of the coherence and 
formation lengths act in opposite directions,
their interplay leads to a nontrivial
$x_F$-dependence of the nuclear transparency as is
shown in fig.~\ref{fig2}.  We present the results
of exact solution of the two-channel problem,
described in $^{9)}$. 

Note that according to Fig.~\ref{fig2} we expect a
decreasing energy-dependence of $S^{pA}_{\Psi}$ as
function of energy at fixed $x_F$.  This may
explain the observed $^{1,2)}$ energy
dependence of nuclear suppression. 
 We remind that our calculations are
restricted to small values of $x_F$.\\

\bigskip

 \noi{\bf 4. Nucleus-nucleus collisions}

\medskip

One may expect new phenomena in heavy ion
collisions.  First of all, the multiparticle
production becomes so intensive that it may cause
an additional suppression of charmonium.  There
might be also an unusual phenomenon, a quark-gluon
plasma formation, in such collisions.  We still
have no reliable calculations of those
effects, but in any case one needs a solid theoretical base
line to compare the measurements with.  The so called
standard absorption model, which corresponds to
$l_c=l_f=0$, is obviously oversimplified, because 
existence of the
quantum interference effects, discussed above, 
 is notnegotiable, and they are very important

\begin{figure}[tbh]
\includegraphics{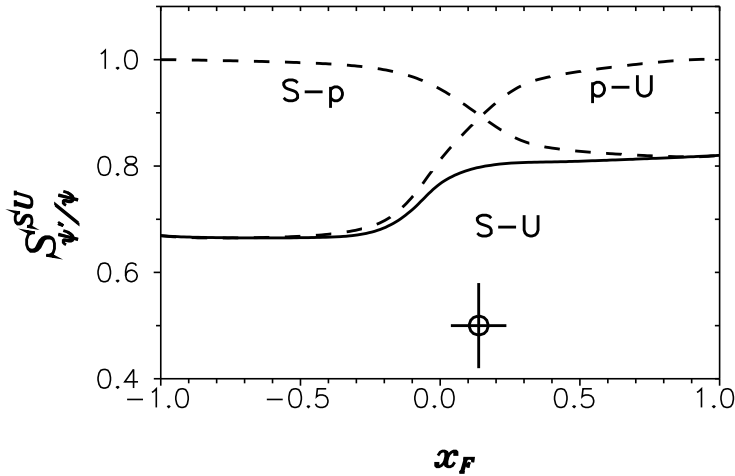}
\begin{center}
\vspace{6cm}
\parbox{13cm}
 {\caption[Delta]
{$x_F$-dependence of relative nuclear suppression
for $\Psi'$ to $\Psi$ in $S-Au$ collisions at
$200\ GeV$.  The solid curve shows prediction of
the two-coupled channel approach $^{5)}$.  The
dashed curves show nuclear suppression in $p-S$
and $p-Au$ collisions.}
\label{fig3}}
\end{center}
\end{figure}

We expect the nuclear effects for charmonium
production in AA collisions to be quite different
from what is known for pA collision.  This is
because the coherence and formation lengths depend on whether
we are in the rest frame of the target or of the
beam.  Due to the inverse kinematics the
charmonium wave packet attenuates with different
effective cross sections propagating through the
two colliding nuclei.  The nuclear suppression in
$A_1A_2$ collision is simply related to that in
$pA_1$ and $pA_2$ interactions,
 \beq
 S^{A_1A_2}_{\Psi}(x_F) \approx
 S^{pA_2}_{\Psi}(x_F)
 S^{pA_1}_{\Psi}(-x_F)
\label{last}
\eeq
Our previous conclusion about equal nuclear
suppression of $\Psi$ and $\Psi'$ in $pA$ interaction
is not valid for $AA$ collisions.
 The results of
application of (\ref{last}) to the ratio
$S^{AB}_{\Psi'}/S^{AB}_{\Psi}$ $^{5)}$ are shown
in Fig.~\ref{fig3}.

We see that this ratio, which is unity in $pA$
interactions is substantially below one in the
case of nuclear collisions.  However, this
reduction explains only about a half of the
observed effect, shown in Fig.~\ref{fig3} by the
only available experimental point $^{3)}$.\\


 \noi{\bf 5. Conclusions}

\medskip

The quantum effects related to interference
between
amplitudes of charmonium production on different
nucleons and to a composite structure of the
produced charmonium wave packet lead to a
substantial modification of the theoretical
expectations. We predict quite an unusual
$x_F$- and energy-dependence of the nuclear
suppression for $\Psi$, which agree with
available data.  We are also able to explain why
the nuclear suppression factors for $\Psi$ and
$\Psi'$ are the same in proton-nucleus, but
different in nucleus-nucleus collisions.
Numerically, however, the observed effect seems to
be larger.  This invites one to take into account
the interaction of the charmonium with other produced
particles, which is the next step to be done.\\

\bigskip

\noi{\bf References}
\baselineskip 20pt

\medskip

\noi
1. D.M.~Adle et al., Phys. Rev. Lett. {\bf 66} (1991)
 133\\
2. The NA3 Collab., J.~Badier et al., Z. Phys. {\bf C 20} (1983)
101\\
3. The NA38 Collaboration, C.~Baglin et al., Phys.
 Lett. {\bf
B345} (1995) 617 and M.C.~Abreu et al., presented at
 QUARK MATTER'95\\
4. S.J.~Brodsky, P.~Hoyer, A.H.~M\"uller and W.-K.~Tang,
Nucl. Phys.
 {\bf B 369} (1992) 519\\
5. J.~H\"ufner and B.Z.~Kopeliovich, Phys. Rev. Lett.
{\bf 76} (1996) 192\\
6. J.~H\"ufner, B.Z.~Kopeliovich and
J.~Nemchik, DOE/ER/40561-260-INT96-19-03,
nucl-th/9605007\\
7. O.~Benhar, B.Z.~Kopeliovich, Ch.~Mariotti, N.N.~Nikolaev
 and
 B.G.~Zakharov, Phys.Rev.Lett. {\bf 69} (1992) 1156.\\
8. B.Z.~Kopeliovich and B.G.~Zakharov, Phys. Rev. {\bf
 D44} (1991) 3466\\
9. J.~H\"ufner and B.Z.~Kopeliovich, paper in preparation

\end{document}